\documentstyle[12pt]{article}

\begin{document}

\newcommand{\wx}{\widetilde{x}}
\newcommand{\wa}{\widetilde{a}}
\newcommand{\wb}{\widetilde{b}}
\newcommand{\wc}{\widetilde{c}}
\newcommand{\wL}{\widetilde{L}}
\newcommand{\wA}{\widetilde{A}}
\newcommand{\wB}{\widetilde{B}}
\newcommand{\wX}{\widetilde{X}}

\newcommand{\cth}{{\rm coth}}
\newcommand{\sh}{{\rm sinh}}

\newcommand{\lag}{\langle}
\newcommand{\rag}{\rangle}

\newcommand{\cM}{{\cal M}}

\begin{titlepage}{\LARGE
\begin{center} Integrable discretizations\\
 of the spin Ruijsenaars--Schneider models
 \end{center}} \vspace{1.5cm}
\begin{flushleft}{\large Orlando RAGNISCO}\end{flushleft} \vspace{0.5cm}
Dipartimento di Fisica, Terza Universita di Roma,\\
Via Vasca Navale 84, Roma, Italy\\
e-mail: ragnisco @ roma1.infn.it

\vspace{0.5cm}

and 

\vspace{0.5cm}

\begin{flushleft}{\large Yuri B. SURIS}\end{flushleft} \vspace{0.5cm}
Centre for Complex Systems and Visualization, University of Bremen,\\
Kitzbuehler Str. 2, 28359 Bremen, Germany\\
e-mail: suris @ cevis.uni-bremen.de

\vspace{1.0cm}

{\small {\bf Abstract.} Integrable discretizations are introduced for the
rational and hyperbolic spin Ruijsenaars--Schneider models. These discrete
dynamical systems are demonstrated to belong to the same integrable
hierarchies as their continuous--time counterparts. Explicit solutions
are obtained for arbitrary flows of the hierarchies, including the
discrete time ones.}
\end{titlepage}

\setcounter{equation}{0}
\section{Introduction}
The famous Calogero--Moser many--particle model possesses the most ramified 
tree of all possible kinds of generalizations. Even at  the 
classical (non--quantum) level, one has three types of the usual Calogero--Moser (CM)
systems: rational, hyperbolic/trigonometric, and elliptic \cite{OP}, \cite{KKS}.
They may be viewed as being related to the $A_{N-1}$ root system, and there
exist immediate generalizations to other root systems \cite{OP}. A relativistic
generalization of the CM systems constitute the Ruijsenaars--Schneider (RS)
models \cite{RS}, \cite{R}, appearing also in three variants  (rational,
hyperbolic/trigonometric, and elliptic). Some time ago spin generalizations
of the CM models were introduced \cite{GH}, \cite{W1}, recently they attracted
more attention \cite{BAB}, \cite{ABB}, \cite{KBBT}. Finally,  a spin 
generalization of the RS model was introduced and studied very recently
\cite{KZ}; it found its application also to the solitons of affine Toda 
theories \cite{BH}. (We have not mentioned some other directions of generalizing
the CM model, such as imposing external potentials etc.)

Still another direction of the development of CM type of models is their 
{\it time-discretization}, which was started by finding the discrete time CM
system in \cite{NP} and followed by the discretization of the RS system
in \cite{NRK}. (Actually, the discretizations of the CM model and some its
generalizations appeared a decade earlier in the image of B\"acklund 
transformations \cite{W2}). The integrable discretizations found in these papers 
turned out to have a remarkable property: they belong to the same integrable 
hierarchies as their continuous time counterparts, i.e. they share with 
continuous time systems such attributes as Hamiltonian structure, Lax matrices, 
$r$--matrices, action--angle variables and so on. This property was clearly 
identified and put in a basis of a general procedure of discretization of 
integrable systems in \cite{S3}, \cite{S4}, for the further development see  
\cite{RB}, \cite{S5}, \cite{S6}.

In the present paper we continue this line of research by applying this general
procedure to the spin RS model. Despite the fact that a Hamiltonian structure
for this recently introduced system is still unknown, it is possible to attach
a hierarchy of flows to this model and to solve explicitly an initial value 
problem for a general flow of the hierarchy. This is done (for the rational and 
hyperbolic systems) in Sect.3. The discrete time systems are introduced and 
studied in Sect.4. In Sect.5 we show what do our results give for the usual RS 
model (without spins): even in this case we obtain new results, or better, give 
new insights. Several open problems are outlined in Sect.6.

\setcounter{equation}{0}
\section{Spin RS model}

We prefer to give here the results for the hyperbolic and rational spin
RS models in a form suitable for our present purposes, since it is
difficult to extract the corresponding results from the general setting of
\cite{KZ}.

The model is described in terms of $N$ scalar variables $x_k$, $1\le k\le N$,
$N$ vectors $|a_k\rangle\in{\bf R}^d$, $1\le k\le N$, and $N$ covectors
$\langle b_k|\in{\bf R}^d$, $1\le k\le N$. We shall represent vectors 
$|a_k\rangle$ as columns of the $d\times N$ matrix $A$, and covectors 
$\langle b_k|$ as rows of the $N\times d$ matrix $B$.

The spin RS model proper as given in \cite{KZ} is described by the equations 
of motion:
\begin{eqnarray}
\dot{x}_k & = & \lag b_k|a_k\rag,\label{dot x}\\
|\dot{a}_k\rag & = & -\lambda_k|a_k\rag
-\sum_{j\neq k}|a_j\rag\lag b_j|a_k\rag\;V(x_j-x_k);\label{dot a}\\
\lag\dot{b}_k| & = & \lambda_k\lag b_k|
+\sum_{j\neq k}\lag b_k|a_j\rag\lag b_j|\;V(x_k-x_j).\label{dot b}
\end{eqnarray}

Here for the rational model
\begin{equation}
V(x)=\frac{1}{x}-\frac{1}{x+\gamma}=\frac{\gamma}{x(x+\gamma)},
\end{equation}
and for the hyperbolic one
\begin{equation}
V(x)=\cth(x)-\cth(x+\gamma)=\frac{\sh(\gamma)}{\sh(x)\sh(x+\gamma)},
\end{equation}
$\gamma$ being parameter of the model. Further, $\lambda_k=\lambda_k(x,a,b)$ in 
(\ref{dot a}), (\ref{dot b}) are arbitrary functions of the dynamical variables.
The simplest possible choice is, of course, $\lambda_k=0$. Solutions of the
system with arbitrary $\lambda_k$'s may be obtained from the solutions of the
system with $\lambda_k=0$ upon change of variables 
$|a_k\rag\mapsto\Lambda_k|a_k\rag$, 
$\lag b_k|\mapsto\Lambda_k^{-1}\lag b_k|$ with 
$\Lambda_k=\exp\Big(\int^t \lambda_k(t')dt'\Big)$.

These models admit  Lax representations
\begin{equation}
\dot{L}=[M,L]
\end{equation}
with the Lax matrix, which we prefer to choose in the form
\begin{equation}\label{L r}
L=L(x,a,b)=\sum_{j,k=1}^N\frac{\gamma}{x_j-x_k+\gamma}\lag b_j|a_k\rag E_{jk},
\end{equation}
or
\begin{equation}\label{L h}
L=L(x,a,b)=\sum_{j,k=1}^N\frac{\sh(\gamma)e^{x_k-x_j}}{\sh(x_j-x_k+\gamma)}
\lag b_j|a_k\rag E_{jk},
\end{equation}
in the rational and hyperbolic cases, respectively. The matrix $M=M(x,a,b)$ has 
the entries
\begin{equation}\label{M0 jk}
M_{jk}=V(x_j-x_k)\lag b_j|a_k\rag,\quad j\neq k,
\end{equation}
\begin{equation}\label{M0 kk}
M_{kk}=\lambda_k.
\end{equation}
In fact, provided the ansatz (\ref{L r}) or (\ref{L h}) is given, the Lax
equation (\ref{L dot}) may be derived as a consequence of the following two
equations, which are in a sense more fundamental:
\begin{equation}
\dot{A}=-AM,\quad \dot{B}=MB.
\end{equation}

So far,  to our knowledge, the Hamiltonian formulation of this model has not been derived. 
This makes it impossible to give a rigorous statement 
about complete integrability of the spin RS model. However, an explicit 
theta--function solution for this model was given in the \cite{KZ}, and we 
shall now demonstrate that in the rational and hyperbolic cases a 
simpler solution can be given, in terms of eigenvalues of certain matrices 
constructed explicitly from the initial conditions, just as for usual CM and RS 
models \cite{OP}, \cite{RS}, and for the spin CM model \cite{GH}. 

\setcounter{equation}{0}
\section{Spin RS hierarchy}

We want to show now that the rational and the hyperbolic spin RS models 
described above are in fact only the simplest representatives of the two corresponding 
hierarchies, and to derive Lax representations for each flow of these 
hierarchies. The members of the hierarchies are encoded by an arbitrary
conjugation--{\it{covariant}} function $f(L)$ on $gl(N)$, the spin RS models proper 
corresponding to the case $f(L)=L$. (Note that in the hypothetical Hamiltonian
formulation the members of the hierarchies would be encoded by 
conjugation--{\it{invariant}} Hamiltonian functions $\varphi(L)$, such that $f(L)=
L\nabla\varphi(L)=\nabla\varphi(L)L$, cf. \cite{S1} in the non--spin case.)

For a conjugation--covariant function $f(L)$ we introduce the system of 
differential equations: 
\begin{eqnarray}
\dot{x}_k & = & f(L)_{kk},\label{x dot}\\
|\dot{a}_k\rag & = & -\lambda_k|a_k\rag-\sum_{j\neq k}|a_j\rag M_{jk},
\label{a dot}\\
\lag\dot{b}_k| & = & \lambda_k\lag b_k|+\sum_{j\neq k}M_{kj}\lag b_j|;
\label{b dot}
\end{eqnarray}
here in the rational case $L$ is defined by (\ref{L r}), and
\begin{equation}\label{M jk r}
M_{jk}=\frac{f(L)_{jk}}{x_j-x_k},\quad j\neq k;
\end{equation}
in the hyperbolic case $L$ is defined by (\ref{L h}), and
\begin{equation}\label{M jk h}
M_{jk}=\frac{e^{x_j-x_k}}{\sh(x_j-x_k)}f(L)_{jk},\quad j\neq k;
\end{equation}
the scalars $\lambda_k$, $1\le k\le N$ are, as before, arbitrary functions
on $x,a,b$. (It is easy to see that we recover the system 
(\ref{dot x})--(\ref{dot b}) for $f(L)=L$).

As the spin RS model proper, this general system also admits Lax representation
\begin{equation}\label{L dot}
\dot{L}=[M,L],
\end{equation}
which, given the ansatz (\ref{L r}) or (\ref{L h}) for the Lax matrix, is a 
consequence of 
\begin{equation}\label{AB dot}
\dot{A}=-AM,\quad \dot{B}=MB.
\end{equation}
Here the off--diagonal entries of the matrix $M$ are given by (\ref{M jk r})
or (\ref{M jk h}), respectively, while the diagonal ones -- by 
\begin{equation}\label{M kk}
M_{kk}=\lambda_k.
\end{equation}

To solve the system explicitly for an arbitrary $f(L)$, we need to introduce
an auxiliary diagonal matrix $X$ by the following formulas: in the rational
case
\begin{equation}\label{X r}
X={\rm diag}(x_1,\ldots,x_N),
\end{equation}
and in the hyperbolic case
\begin{equation}\label{X h}
X={\rm diag}(e^{2x_1},\ldots,e^{2x_N}).
\end{equation}

It is important to note that the ansatz (\ref{L r}) for the rational Lax
matrix is equivalent to
\begin{equation}\label{mom r}
XL-LX+\gamma L=\gamma BA,
\end{equation}
while the ansatz (\ref{L h}) for the hyperbolic Lax matrix is equivalent to
\begin{equation}\label{mom h}
e^{\gamma}XLX^{-1}-e^{-\gamma}L=2\sh(\gamma)BA.
\end{equation}

A key observation necessary for an explicit integration of equations of motion, 
is the following: the evolution equation for $x_k$ (\ref{x dot}) together with 
the expressions (\ref{M jk r}), (\ref{M jk h}) may be cast in a single  matrix 
differential equation for $X$, whose form  depends on the case under scrutiny. Namely, for 
the rational one:
\begin{equation}\label{X dot r}
\dot{X}=[M,X]+f(L),
\end{equation}
and for the hyperbolic one:
\begin{equation}\label{X dot h}
\dot{X}=[M,X]+2Xf(L).
\end{equation}

Now we proceed as in \cite{OP}, \cite{RS}, \cite{GH}: define the matrix
$V=V(t)$ as the solution of the matrix differential equation
\begin{equation}
\dot{V}V^{-1}=M
\end{equation}
with the initial condition $V(0)=I$. Then in the rational case
\begin{equation}\label{X(t) r}
X=X(t)=V(X_0+tf(L_0))V^{-1},
\end{equation}
in the hyperbolic case
\begin{equation}\label{X(t) h}
X=X(t)=VX_0e^{2tf(L_0)}V^{-1},
\end{equation}
and in the both cases
\begin{equation}\label{L(t)}
L=L(t)=VL_0V^{-1}. 
\end{equation}
\begin{equation}\label{AB(t)}
A=A(t)=A_0V^{-1},\quad B=B(t)=VB_0.
\end{equation}

So, just as in the case of the usual CM and RS models and the spin CM model, 
in the rational case the positions $x_k$ are the eigenvalues of the matrix
$X_0+tf(L_0)$, and in the hyperbolic case $\exp(2x_k)$ are the eigenvalues
of the matrix $X_0\exp(2tf(L_0))$. As in the spin CM model, the corresponding
eigenvectors (i.e. the diagonalizing matrix $V$) define the evolution of spin
degrees of freedom. The non--uniqueness of the diagonalizing matrix (which is
in principle defined up to a left multiplication by a diagonal matrix) is
reflected in the arbitrariness of the diagonal part of the matrix 
$M=\dot{V}V^{-1}$, i.e. of the scalars $\lambda_k$ (cf. (\ref{M kk})).

\setcounter{equation}{0}
\section{Integrable discretizations of the spin RS models}

We introduce now two families of maps 
\[
(x_k,|a_k\rag,\lag b_k|) \mapsto (\wx_k,|\wa_k\rag,\lag \wb_k|),
\]
depending on the small parameter $h>0$ and approximating, as $h\to 0$, the
rational and hyperbolic spin RS models, respectively. Just as the spin RS models
have $N$ free parameters $\lambda_k=\lambda_k(x,a,b)$, will our maps have
$N$ free parameters $\mu_k$.

The rational spin RS map is defined by:
\begin{equation}\label{wx r}
\wx_k-x_k=h\mu_k^{-1}\lag \wb_k|a_k\rag,
\end{equation}
\begin{equation}\label{wa r}
\mu_k|\wa_k\rag-|a_k\rag=
-h\left(\sum_{j\neq k}\frac{|\wa_j\rag\lag \wb_j|a_k\rag}{\wx_j-x_k}-
\sum_{j=1}^N \frac{|a_j\rag\lag b_j|a_k\rag}{x_j-x_k+\gamma}\right),
\end{equation}
\begin{equation}\label{wb r}
\lag\wb_k|-\mu_k\lag b_k|=
h\left(\sum_{j\neq k}\frac{\lag \wb_k|a_j\rag\lag b_j|}{\wx_k-x_j}-
\sum_{j=1}^N \frac{\lag \wb_k|\wa_j\rag\lag\wb_j|}{\wx_k-\wx_j+\gamma}\right),
\end{equation}
where the parameters $\mu_k$ are supposed to satisfy the following asymptotic 
relations:
\begin{equation}\label{asym mu}
\mu_k(x,\wx,a,\wa,b,\wb;h)=1+h\mu^{(0)}_k(x,\wx,a,\wa,b,\wb)+O(h^2).
\end{equation}
The simplest possible choice is, of course, $\mu_k=1$.

The standard implicit functions theorem implies that the equations
(\ref{wx r})--(\ref{wb r}) for $h$ small enough  have a unique solution
$(\wx,\wa,\wb)$ which is $O(h)$--close to $(x,a,b)$. Hence a map
$(x,a,b)\mapsto(\wx,\wa,\wb)$ is defined, being an approximation of the  
time $h$ shift along the trajectories of the rational version of the system 
(\ref{dot x})--(\ref{dot b}) with
\begin{equation}
\lambda_k(x,a,b)=\mu_k^{(0)}(x,x,a,a,b,b)-\gamma^{-1}\lag b_k|a_k\rag
\end{equation}
(the second term in the right--hand side of this formula stems from the fact 
that in the second sums in (\ref{wa r}), (\ref{wb r}) the term corresponding
to $j=k$ is present, as opposed to (\ref{dot a}), (\ref{dot b})). The choice 
$\mu_k=1$ corresponds,  therefore, to the flow of the rational spin RS model 
with $\lambda_k=-\gamma^{-1}\lag b_k|a_k\rag$. In order to get an approximation 
for the flow with $\lambda_k=0$ one can take, for example, 
$\mu_k=1+h\gamma^{-1}\lag\wb_k|a_k\rag$.

The hyperbolic spin RS map is defined by:
\begin{eqnarray}
e^{2(\wx_k-x_k)}-1 & = & 2h\mu_k^{-1}\lag \wb_k|a_k\rag,
\label{wx h}\\
\mu_k|\wa_k\rag-|a_k\rag     & = &
-h\bigg[\sum_{j\neq k}|\wa_j\rag\lag \wb_j|a_k\rag\Big(\cth(\wx_j-x_k)-1\Big)
\nonumber\\
                             &   & 
-\sum_{j=1}^N |a_j\rag\lag b_j|a_k\rag\Big(\cth(x_j-x_k+\gamma)-1\Big)\bigg],
\label{wa h}\\
\lag\wb_k|-\mu_k\lag b_k|    & = &
h\bigg[\sum_{j\neq k}\lag \wb_k|a_j\rag\lag b_j|\Big(\cth(\wx_k-x_j)-1\Big)
\nonumber\\
                             &   &
-\sum_{j=1}^N\lag\wb_k|\wa_j\rag\lag\wb_j|\Big(\cth(\wx_k-\wx_j+\gamma)-1\Big)
\bigg].
\label{wb h}
\end{eqnarray}
Under the same assumtion (\ref{asym mu}) as before, this system defines a map
$(x,a,b) \mapsto (\wx,\wa,\wb)$ approximating the time $h$ shift along the
trajectories of the hyperbolic system (\ref{dot x})--(\ref{dot b}) with
\begin{equation}
\lambda_k(x,a,b)=\mu_k^{(0)}(x,x,a,a,b,b)-(\cth(\gamma)-1)\lag b_k|a_k\rag.
\end{equation}
The choice $\mu_k=1$ corresponds this time to the flow of the hyperbolic spin
RS model with $\lambda_k=-(\cth(\gamma)-1)\lag b_k|a_k\rag$.

An outstanding property of the introduced discretizations for the spin RS models
is their solvability. More precisely, they turn out to belong to the same
hierarchies as their continuous time counterparts. That means that they admit
discrete time Lax representations with the same Lax matrices as the 
continuous systems. 

{\bf Theorem 1.} {\it The maps} (\ref{wx r})--(\ref{wb r}) {\it and}
(\ref{wx h})--(\ref{wb h}) {\it have the following discrete Lax representations:
\begin{equation}\label{d L}
\widetilde{L}\cM=\cM L
\end{equation}
\begin{equation}\label{d AB}
\wA\cM=A(I+h\alpha L),\quad  (I+h\alpha\wL)\wB=\cM B.
\end{equation}
Here for the rational case $\alpha=\gamma^{-1}$, $L$ is as in} (\ref{L r}),
{\it and
\begin{equation}\label{cM r}
\cM_{jk}=\frac{h\lag \wb_j|a_k\rag}{\wx_j-x_k};
\end{equation}
for the hyperbolic case $\alpha=\cth(\gamma)-1$, $L$ is as in} (\ref{L h}), 
{\it and
\begin{equation}\label{cM h}
\cM_{jk}=\frac{he^{x_k-\wx_j}}{\sh(\wx_j-x_k)}\lag \wb_j|a_k\rag;
\end{equation}
in the both cases the condition
\begin{equation}
\cM_{kk}=\mu_k
\end{equation}
is imposed.}

The {\bf proof} of this Theorem follows by direct calculation. It should
be remarked that the same phenomenon as in the continuous cases takes place,
namely that the evolution equations (\ref{d AB}) for $A$, $B$ are in a sense 
more fundamental then the Lax equation (\ref{d L}) for $L$: given the ansatz
(\ref{L r}) or (\ref{L h}) for $L$ and a corresponding ansatz (\ref{cM r}) or
(\ref{cM h}) for $\cM$, the Lax equation (\ref{d L}) {\it is a consequence}
of (\ref{d AB}).

From the Lax equation (\ref{d L}) follows that the equations (\ref{d AB})
can be presented also as 
\begin{equation}\label{d AB mod}
\wA=A(I+h\alpha L)\cM^{-1},\quad  \wB=\cM(I+h\alpha L)^{-1}B.
\end{equation}
As a consequence we get: $\wA\wB=AB$, or
\[
\sum_{j=1}^N|\wa_j\rag\lag \wb_j|=\sum_{j=1}^N|a_j\rag\lag b_j|.
\]
This allows to rewrite the hyperbolic map (\ref{wx h})--(\ref{wb h}) in the 
following form:
\begin{eqnarray}
{\rm tanh}(\wx_k-x_k) & = & h\nu_k^{-1}\lag \wb_k|a_k\rag,
\label{wx h mod}\\
\nu_k|\wa_k\rag-|a_k\rag     & = &
-h\bigg(\sum_{j\neq k}|\wa_j\rag\lag \wb_j|a_k\rag\cth(\wx_j-x_k)
\nonumber\\
                             &   & 
-\sum_{j=1}^N |a_j\rag\lag b_j|a_k\rag\cth(x_j-x_k+\gamma)\bigg),
\label{wa h mod}\\
\lag\wb_k|-\nu_k\lag b_k|    & = &
h\bigg(\sum_{j\neq k}\lag \wb_k|a_j\rag\lag b_j|\cth(\wx_k-x_j)
\nonumber\\
                             &   &
-\sum_{j=1}^N\lag\wb_k|\wa_j\rag\lag\wb_j|\cth(\wx_k-\wx_j+\gamma)\bigg)
\label{wb h mod}
\end{eqnarray}
(where $\nu_k=\mu_k+h\lag\wb_k|a_k\rag$, so that, for example, the choice
$\nu_k=1$ leads to a discretization of the hyperbolic spin RS model with
$\lambda_k=-\cth(\gamma)\lag b_k|a_k\rag$).

It turns out to be possible to find continuous time flows from
the spin RS hierarchies {\it interpolating} our maps (we say that a map is
interpolated by a flow, if each orbit of the map is given by the values at
$t=nh$, $n\in{\bf Z}$, of a certain orbit of the flow). The following statement
gives these interpolating flows. This, further, provides us with explicit 
solutions of the introduced discrete time dynamical systems.

{\bf Theorem 2.} {\it The map} (\ref{wx r})--(\ref{wb r}) {\it is interpolated 
by the flow of the rational spin RS hierarchy} (\ref{x dot})--(\ref{b dot})
{\it with the function
\[
f(L)=L(I+h\gamma^{-1}L)^{-1}.
\]
Analogously, the map} (\ref{wx h})--(\ref{wb h}) {\it is interpolated 
by the flow of the hyperbolic spin RS hierarchy} (\ref{x dot})--(\ref{b dot})
{\it with the function}
\[
f(L)=\frac{1}{2h}
\log\left(\frac{I+h(\cth(\gamma)+1)L}{I+h(\cth(\gamma)-1)L}\right).
\]

{\bf Proof for the rational case.} Equations of motion (\ref{d L}), 
(\ref{d AB mod}), where $\alpha=\gamma^{-1}$, have to be supplemented with an 
ansatz for $\cM$, which may be also represented as an equation of motion 
for the auxiliary diagonal matrix $X$ from (\ref{X r}):
\[
\widetilde{X}\cM-\cM X=h\widetilde{B}A.
\]
Introduce now the matrix 
\[
M=\cM(I+h\gamma^{-1}L)^{-1}.
\]
In terms of this matrix evolution equations may be presented as
\begin{equation}\label{disc L}
\wL=MLM^{-1},
\end{equation}
\begin{equation}\label{disc AB}
\wA=AM^{-1},\quad \wB=MB,
\end{equation}
\begin{equation}\label{disc X}
\wX M(I+h\gamma^{-1}L)-M(I+h\gamma^{-1}L)X=hMBA.
\end{equation}
But the ansatz (\ref{L r}) for $L$, being equivalent to the commutation relation
(\ref{mom r}), implies:
\[
(I+h\gamma^{-1}L)X=X(I+h\gamma^{-1}L)+hL-hBA.
\]
This allows to rewrite (\ref{disc X}) in the form
\begin{equation}\label{discr X}
\wX M-MX=hML(I+h\gamma^{-1}L)^{-1}=hMf(L).
\end{equation}
Defining now $V$ as the solution of the matrix difference equation
\begin{equation}\label{wV}
\widetilde{V}V^{-1}=M
\end{equation}
with an initial condition $V_0=I$, we see immediately that the solution of 
(\ref{disc L}), (\ref{disc AB}), (\ref{discr X}) is given by
\begin{equation}\label{LABn}
L=VL_0V^{-1},\quad A=A_0V^{-1},\quad B=VB_0,
\end{equation}
\begin{equation}\label{Xn r}
X=V(X_0+nhf(L_0))V^{-1},
\end{equation}
which proves the Theorem in the rational case.

{\bf Proof for the hyperbolic case.} The equations of motion (\ref{d L}),
(\ref{d AB mod}), where this time $\alpha=e^{-\gamma}/\sh(\gamma)$, have to be 
supplemented with an ansatz for $\cM$, which is this time equivalent to the 
following equation of motion for the auxiliary diagonal matrix $X$ from 
(\ref{X h}):
\[
\widetilde{X}\cM-\cM X=2h\widetilde{B}AX.
\]
In terms of the matrix 
\[
M=\cM(I+h\alpha L)^{-1}
\]
the evolution equations for $L$, $A$, $B$ take the form (\ref{disc L}),
(\ref{disc AB}), and for $X$ -- the form
\begin{equation}\label{disc X h}
\wX M(I+h\alpha L)-M(I+h\alpha L)X=2hMBAX.
\end{equation}
The ansatz (\ref{L h}) for $L$ is equivalent to the commutation relation
(\ref{mom h}), which, in turn, implies:
\[
(I+h\alpha L)X=X(I+h\alpha e^{2\gamma}L)-2hBAX
\]
(where we have taken into account that $\alpha e^{\gamma}\sh(\gamma)=1$).
This allows to rewrite (\ref{disc X h}) in the form
\begin{equation}\label{discr X h}
\wX M=MX\frac{I+h\alpha e^{2\gamma}L}{I+h\alpha L}=MXe^{2hf(L)}.
\end{equation}
Defining now $V$ by (\ref{wV}), we see immediately that the solution of 
(\ref{disc L}), (\ref{disc AB}), (\ref{discr X h}) is given by (\ref{LABn}) and
\begin{equation}\label{Xn h}
X=VX_0\left(\frac{I+h\alpha e^{2\gamma}L_0}{I+h\alpha L_0}\right)^nV^{-1}=
VX_0e^{2nhf(L_0)}V^{-1},
\end{equation}
which proves the Theorem in the hyperbolic case.

\setcounter{equation}{0}
\section{Connection with the usual RS model}

In the case $d=1$, i.e when $|a_k\rag$, $\lag b_k|$ become scalars, the
spin RS models yield back the usual spinless ones \cite{RS}, \cite{R}. The usual RS
hierarchy is now well studied. In particular, explicit solutions were found in 
\cite{RS}, \cite{R} for the rational and hyperbolic cases, and in \cite{KZ} in 
the elliptic one; the Hamiltonian structure of the usual RS hierarchy is known 
and admits an $r$--matrix interpretation \cite{AR}, \cite{S1}, \cite{NKSR}, 
\cite{S2}. 

To obtain the usual RS model from its spin counterpart in the case $d=1$, one 
sets $a_kb_k=c_k$ in (\ref{x dot})--(\ref{b dot}), and it follows {\it 
irrespective} of the values of the parameters $\lambda_k$:
\[
\dot{x}_k=c_k,
\]
\[
\dot{c}_k=c_k\sum_{j\neq k}c_j\Big[V(x_k-x_j)-V(x_j-x_k)\Big].
\]
These equations describe the usual RS model, the simplest flow of the usual RS
hierarchy.

The same reduction is admissible for our discrete systems. Namely, {\it
irrespective} of the values of parameters $\mu_k$ the rational equations
(\ref{wx r})--(\ref{wb r}) imply:
\begin{equation}\label{1 r}
h\sum_{j=1}^N\frac{\wc_j}{\wx_j-x_k}
-h\sum_{j=1}^N\frac{c_j}{x_j-x_k+\gamma} = 1,
\end{equation}
\begin{equation}\label{2 r}
-h\sum_{j=1}^N\frac{\wc_j}{\wx_k-\wx_j+\gamma}
+h\sum_{j=1}^N\frac{c_j}{\wx_k-x_j} = 1,
\end{equation}
and the hyperbolic equations (\ref{wx h mod})--(\ref{wb h mod}) imply:
\begin{equation}\label{1 h}
h\sum_{j=1}^N\wc_j\cth(\wx_j-x_k)-h\sum_{j=1}^Nc_j\cth(x_j-x_k+\gamma) = 1,
\end{equation}
\begin{equation}\label{2 h}
-h\sum_{j=1}^N\wc_j\cth(\wx_k-\wx_j+\gamma)+h\sum_{j=1}^Nc_j\cth(\wx_k-x_j)=1.
\end{equation}

These equations are very similar to those found in \cite{NRK}, and can be 
analysed along the same lines. Namely, each pair of equations (\ref{1 r}),
(\ref{2 r}) or (\ref{1 h}), (\ref{2 h}) may be considered as a linear system
for $2N$ variables $c_k$, $\wc_k$ with a Cauchy matrix. This system may be
explicitly solved, which allows then to write down ''Newtonian'' equations
of motion in terms of $x$ variables only.

To appreciate the difference between the equations found in \cite{NRK} and here,
it will be instructive to compare the Lax representations for them. Our results
immediately imply the Lax representation of the form
\[
\wL\cM=\cM L
\]
where in the rational case
\[
L=\sum_{j,k=1}^N\frac{\gamma}{x_j-x_k+\gamma}c_kE_{jk},\quad
\cM=\sum_{j,k=1}^N\frac{h}{\wx_j-x_k}c_kE_{jk},
\]
and in the hyperbolic case
\[
L=\sum_{j,k=1}^N\frac{\sh(\gamma)e^{x_k-x_j}}{\sh(x_j-x_k+\gamma)}c_kE_{jk},
\quad
\cM=\sum_{j,k=1}^N\frac{he^{x_k-\wx_j}}{\sh(\wx_j-x_k)}c_kE_{jk}.
\]
So, they belong to the same hierarchies as their continuous time counterparts;
the interpolating flows are given by the same formulas as in  Theorem 2.

The Lax representations for the systems found in \cite{NRK} have the same (up
to a simple gauge transformation) Lax
matrices $L$, but different matrices $\cM$, namely in the rational case
\[
\cM=\sum_{j,k=1}^N\frac{\gamma}{\wx_j-x_k+\gamma}c_kE_{jk},
\]
and in the hyperbolic case
\[
\cM=\sum_{j,k=1}^N\frac{\sh(\gamma)e^{x_k-\wx_j}}
{\sh(\wx_j-x_k+\gamma)}c_kE_{jk}.
\]
Hence, they also belong to the same hierarchies as their continuous time 
counterparts, but are interpolated by different flows. In fact, interpolating
flows were found in \cite{NRK} to correspond to
\[
f(L)=-(L+h\gamma^{-1})^{-1}
\]
in the rational case, and to
\[
f(L)=\frac{1}{2h}\log\left(\frac{L+h(\cth(\gamma)-1)I}
{L+h(\cth(\gamma)+1)I}\right)
\]
in the hyperbolic case.

We see  that the maps found in the present Letter serve as approximations to
the flow of the RS hierarchy with $f(L)=L$, which corresponds to the Hamiltonian
function $\varphi(L)={\rm tr}(L)$, while the maps found in \cite{NRK} serve as
approximations to another flow of the RS hierarchy, characterized by  $f(L)=-L^{-1}$, which
corresponds to the Hamiltonian function $\varphi(L)={\rm tr}(L^{-1})$. 
\footnote{{\it Both} maps are derived in the recent preprint \cite{KLWZ}
along with their Lax representations (in the general elliptic case). However,
Hamiltonian interpretation and relations to continuous time hierarchies are
not discussed there.}

\section{Conclusion}
This paper contains the application of a general procedure of integrable 
discretizations to the rational and hyperbolic spin RS models. There arise
several natural questions.
\begin{itemize}
\item {\it What about elliptic spin RS model?} We have not pursued this point,
basically because of the lacking Hamiltonian structure of this model. Indeed,
a natural extension of our results to the elliptic case leads to a discrete Lax 
equation with a spectral parameter $\lambda$:
\[
\wL(\lambda)\cM(\lambda)=\cM(\lambda)L(\lambda)
\]
with the same Lax matrix 
\[
L(\lambda)=\sum_{j,k=1}^N\Phi(x_j-x_k+\gamma,\lambda)\lag b_j|a_k\rag E_{jk}
\]
as for the continuous system, and
\[
\cM(\lambda)=\sum_{j,k=1}^N\Phi(\wx_j-x_k,\lambda)\lag \wb_j|a_k\rag E_{jk}
\] 
(here $\Phi(x,\lambda)=\sigma(x+\lambda)/\sigma(x)\sigma(\lambda)$, and
$\sigma(x)$ is the Weierstrass sigma function). However, in absence of the 
underlying Hamiltonian structure we can not make any assertions about complete 
integrability of the resulting system. A surrogate for this -- an explicit 
solution of arbitrary flow in a hierarchy -- is also not available in the 
elliptic case. We intend to return to the elliptic case in the future.
\item {\it What about spin CM models?} These models, at least their rational
and hyperbolic versions, can be discretized along the same lines as the spin
RS models. However, this (paradoxically) turns out to be a technically more
difficult problem. A reference to the standard implicit function theorem
when discussing the equations of the type (\ref{wx r})--(\ref{wb r}) has to be 
replaced by less trivial algebraic considerations. This will be discussed 
elsewhere.
\end{itemize}

\section{Acknowledgement}
The research of Yu.Suris is financially supported by DFG (Deutsche
Forshungsgemeinschaft). He also thanks the University of Rome for support
of the visit to Rome, where this work was done.

\end{document}